\documentclass[mathleft
]{an}
\usepackage{graphicx}
\begin{document}

\Pagespan{00}{}
\Yearpublication{2010}%
\Yearsubmission{2010}%
\Month{...}%
\Volume{...}%
\Issue{...}%

\title{The EXOTIME project:\\
 a status report on PG 1325+101 (QQ Vir)}

\author{S. Benatti\inst{1}\fnmsep\thanks{Corresponding author:
  \email{serena.benatti@unipd.it}\newline}
\and  R. Silvotti\inst{2}
\and  R. U. Claudi\inst{3}
\and  S. Schuh\inst{4,5}
\and  R. Lutz\inst{4,6}
\and  S.-L. Kim\inst{7}
\and  R. Janulis\inst{8}
\and  M. Papar\`o\inst{9}
\and  A. Baran\inst{10}
\and  R. {\O}stensen\inst{11}
}
\titlerunning{The EXOTIME project: a status report on PG 1325+101 (QQ Vir)}
\authorrunning{S. Benatti et al.}
\institute{
CISAS, Universit\`a degli Studi di Padova, Italy
\and 
INAF Osservatorio Astronomico di Torino, Pino Torinese, Italy
\and 
INAF Osservatorio Astronomico di Padova, Italy
\and
Georg-August-Universit\"at G\"ottingen Institut f\"ur Astrophysik, Germany 
\and
Eberhard-Karls-Universit\"at T\"ubingen Kepler Center / Institut f\"ur Astronomie und Astrophysik, T\"ubingen, Germany 
\and
Max Planck Institut fur Sonnensystemforschung, Katlenburg-Lindau, Germany
\and
Korea Astronomy and Space Science Institute, Daejeon, Korea
\and
Institute of Theoretical Physics and Astronomy, Vilnius University, Lithuania
\and
Konkoly Observatory, Hungary
\and
Cracow Pedagogical University, Poland
\and
Institute for Astronomy, K. U. Leuven, Belgium
}

\received{}
\accepted{}
\publonline{later}


\abstract{  
After the discovery of V391 Peg b, the first planet detected around a post Red Giant phase star (Silvotti et al. 2007), the EXOTIME (EXOplanet search with the TIming MEthod) project is focused on the search for new planets with similar characteristics. The aim of the project is to organize a global observing network to collect as much data as possible for a sample of five subdwarf B (sdB) stars and share them in order to obtain a more precise analysis. These evolved pulsators may have extremely regular oscillation periods. This feature makes these stars suitable to search for planetary companions with the timing method as in the case of pulsars. In this contribution we present the project and some preliminary results for the
star PG 1325+101 (QQ Vir) after the first two years of activity.
}

\maketitle

\section{Introduction}
The EXOTIME project (EXOplanet search with the TIming MEthod) is a coordinated observing program aimed at the search for substellar companions around pulsating subdwarf B stars and the derivation of evolutionary timescales. 
The required observations span over a very long time (of the order of years), so this method is quite expensive in terms of observing time, but with the advantage to be sensitive to wide orbits and relatively low masses (down to $\sim $ M$_{Jup}$). 

Using the timing method we can measure both the variations of the pulsation periods and the phase variations; the latter potentially allow us to detect the presence of substellar companions as in the case of V391 Peg b (Silvotti et al. 2007), the first planet discovered with this technique around a pulsating star.
The discovery of V391 Peg b has raised the interest to investigate evolved planetary systems beyond the main sequence and beyond the red giant branch.
The orbital distance of this planet, lower than 2 AU, suggests that this planet may have ``survived'' to the RG expansion of the parent star.
Recently, two substellar companions have been detected orbiting the sdB eclipsing binary HW Vir (Lee et al. 2009), suggesting that substellar objects might be a relatively common phenomenon around sdB stars.

Further goals of EXOTIME are the characterization of the targets using asteroseismic methods and the measurement of the secular variation of the oscillation period (the so called P-dot, $\dot{P}$), giving a further example of the synergy between asteroseismology and the search for exoplanets, already known in the case of solar-like stars. 
Measuring $\dot{P}$ allows a precise determination of the evolutionary status of a star and can help the identification of the pulsation modes.

As a final goal, EXOTIME wishes to improve our understanding of the formation and evolution of the sdB stars. 
The formation processes of sdB stars still represent an unclear topic in stellar evolution. Different scenarios have been proposed, both for single stars and binaries. The presence of a secondary body such as a planet, in particular for single stars, has been suggested to play a role in this process, enhancing the mass loss near the RGB tip (Soker 1998).

\section{The Method}
As in the case of the pulsar timing, we use the oscillation periods as a clock in order to detect all the possible variations in the travel time of the photons. These variations could be attributed to the secular variation of the oscillation period, $\dot{P}$, or to a wobble of the sdB location due to the presence of a perturbing body, such as a planet. These variations are easily detectable through the O-C diagram, in which the theoretical ({\itshape Calculated}) expectations of a particular quantity are compared with the {\itshape Observed} ones.
Once one obtains a sufficient number of data, it is possible to compare the mean phase of a long monitoring period with the phases at different time steps. If changes occur in the pulsation period, they can be detected and identified according to the distribution of the points in the graph. When a pulsation period changes linearly in time, the O-C diagram shows a parabolic shape caused by the evolutionary timescale. The presence of a companion is revealed by a sinusoidal trend which means cyclically advanced or delayed timings of the maxima (or minima), due to the motion around common barycenter. 

The drawback of timing is the need of long-term monitoring, which imposes to collect data for many years. Anyway this technique is somewhat complementary to the radial velocity and transit methods, since these methods can hardly provide information on stars with small radii and hardly detect planets in wide orbits and relatively low masses (down to $\sim $ M$_{Jup}$). 



\section{Targets}
The targets of EXOTIME (see Table \ref{tlab}) were selected following different criteria.
We excluded known binaries and stars with an IR excess from the 2MASS data.
Other important features are stable amplitude spectra in terms of phases and amplitudes and simple spectra, ideally with only 2 or 3 frequencies. A low number of frequencies allows to resolve the frequency spectrum even in short runs while, on the same time, it gives the opportunity to obtain independent O-C plots from each individual frequency.
Finally we gave higher piority to bright targets, high pulsation amplitudes and high observability, considering that the available telescopes are mostly in the northern emisphere.
\begin{table}[htbp]
 \centering
\caption{Targets of the EXOTIME program. References: (a) {\O}stensen et al. 2001b; (b) Dreizler et al. 2002; (c) Kilkenny et al. 2006; (d) Silvotti et al. 2002; (e) {\O}stensen et al. 2001a. $^{\star}$ Hybrid pulsators.}
\label{tlab}
\begin{tabular}{|l|c|c|c|c|}\hline
Object name & ref. & B & Main & Amplitude \\
 &  & mag &  Period [s] & [mma]  \\
\hline
HS 0444+0458 & a &  15.2 & $\sim$137 & $\sim$11 \\
(V1636 Ori) & & &  & \\ 
\hline
HS 0702+6043 & b&  14.7 & $\sim$360 & $\sim$22 \\
(DW Lyn)$^{\star}$ &  &  &  & \\ 
\hline
EC 09582-1137 & c& 15 & $\sim$136 & $\sim$8\\ 
 &  &  &  $\sim$151 & $\sim$7\\
\hline
PG 1325+101 &d & 13.8  & $\sim$138  & $\sim$26 \\
(QQ Vir) & & & & \\ 
\hline
HS 2201+2610 & e& 14.3& $\sim$350 & $\sim$10 \\
(V391 Peg)$^{\star}$ & &  &  & \\
\hline
\end{tabular}
\end{table}

\section{Observations}
Since 2008 the EXOTIME collaboration collected a large amount of data from several observing sites covering the longitude range between Eastern Europe to Western North America, equipped with telescopes having typical apertures of 1-2 meters (up to 4 m).
The observations are performed in B filter and, according to the pulsation period of the target and the magnitude, the time sampling is set to obtain at least 5-6 points per cycle.
Because of the different characteristics of the instruments, it is necessary to weight the incoming data in order to obtain a coherent data set. 

The list of the observations performed during the first two years of EXOTIME is available from a dedicated web site ({\itshape www.na.astro.it/$\sim$silvotti/exotime/}).
The web site provides informations on the target stars, observing instructions and scheduled observations, useful to optimize the planning of the observations, and the coverage of the targets.


\section{Status of the project}
The EXOTIME program carries on with its activity of collecting data for each target star. Preliminary analysis of HS 0702+6043 and HS 0444+0458 are available in Schuh et al. 2010, while an updated phase analysis including the new data on V391 Peg will be performed in the next months.
The data set collected up to now on EC 09582-1137 is still too poor for our goals. Here we present the preliminary analysis of PG 1325+101 (QQ Vir) using part of the available data. 
Since we know that many of the sdB stars show amplitude variations (and sometimes phase variations) on various time scales from years to days, a preliminary check on the stability of the oscillations is required.

\subsection{PG 1325+101: preliminary analysis} 
As a first step we can exclude the presence of a cooler stellar companion, studying the Spectral Energy Distribution. 
\begin{figure}
\includegraphics[width=80mm]{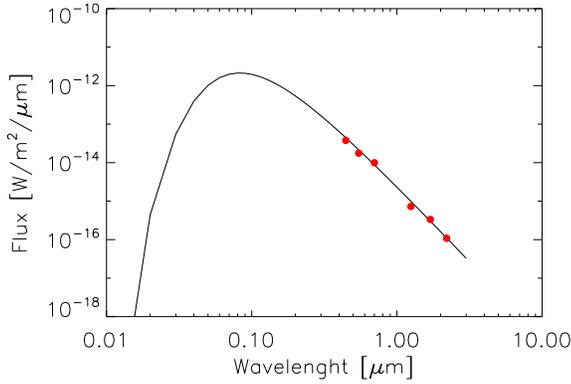}
\caption{Spectral Energy Distribution of PG 1325+101. The filled red circles are 2MASS JHK and BVR magnitudes from the literature. There is no evidence for a cool stellar companion to the sdB.}
\label{sed}
\end{figure}
Figure \ref{sed} shows the comparison between a black body (which provides a good approximation of the energy distribution of a sdB star) and the 2MASS JHK and BVR magnitudes. The plot doesn't show any infrared excess. 

After this check we can start to analyse the incoming data set for this target. Generally the data are reduced directly by the observers,
but since the outputs can be expressed in different ways, it is very important to turn them into a standard format. The time series are then barycentrically corrected and the differential photometric flux is expressed in mmi (milli-modulation
intensity units, 1 mmi=0.1\%). Then we calculate the uncertainty for each data point (following the approach of Silvotti et al. 2006) in order to fix the weights to the different data sets.
The available data are shown in Table \ref{observations} and they sample the seasons 2008 and 2009 from January to June.
\begin{table}[htbp]
 \centering
\caption{Available data for PG 1325+101.}
\label{observations}
\begin{tabular}{|l|l|c|c|}\hline
Date & Telescope & Filter & Lenght\\
\hline
2008/03/01 & 1.5m-Loiano & B & 5.4 h \\
\hline
2008/03/13 & 1m-Piszk\'{e}stet\~{o} & B & 4.8 h\\ 
\hline
2008/02-05 & 1.2 Mercator + & V & 95.5 h \\
 & Suhora \& Baker &   &   \\
\hline
2008/06/04,06 & 1.6m-Moletai & B & 3 h\\ 
\hline
2009/01/28-31 & 1m-LOAO & B & 9 h \\
\hline
2009/02/25-27 & 1.2m-MONET & B & 11.7 h \\
\hline
2009/03/18 & 1.2m-MONET & B & 1.7 h\\
\hline
2009/03/20-22 & 2.2m-Calar Alto & B & 11.5 h \\
\hline
2009/03/25-28 & 1m-LOAO & B & 23.8 h \\
2009/03/30-31 & & & \\
\hline
2009/04/22 & 1.2m-Loiano & B & 2.9 h \\
\hline
2009/04/29 & 2.2m-Calar Alto & B & 2.9 h \\
\hline
2009/05/06 & 2.2m-Calar Alto & B & 3.9 h \\
\hline
2009/05/25-27 & 1m-LOAO & B & 14 h \\
\hline
\end{tabular}
\end{table}

The pulsation frequencies of PG 1325+101 are well known from Silvotti et al. 2006 and Baran et al. 2010.
We can thus fix the frequencies in order to calculate the mean phase of the whole data set using the software Period04 (Lenz and Breger 2005): this represents the {\itshape Calculated} value in the O-C diagram. We repeat the procedure also for the single runs, measuring the {\itshape Observed} values. The phase differences are then turned into time-lag and plotted against the observations periods. At this stage of the analysis we prefer to perform the calculations using only the main pulsation frequency (at 7255.5 $\mu Hz$, with an amplitude of about 27 mma), in order to avoid confusion on the sinusoidal fit and on the estimate of the phase with P04 for the runs with poor frequency resolution.

\begin{figure}
\includegraphics[width=80mm]{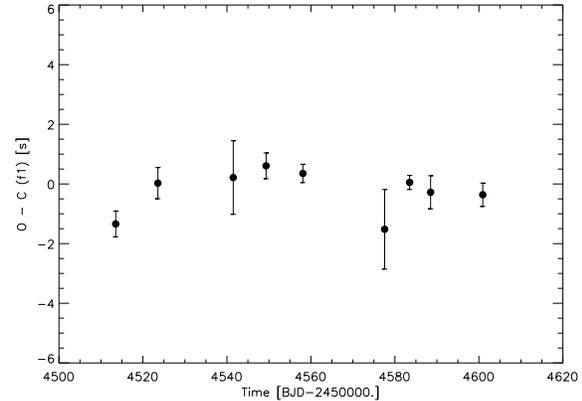}
\caption{O-C diagram of the main frequency (f1) of PG 1325+101 for the V band run.}
\label{vband}
\end{figure}

We have calculated the O-C diagram mantaining separated the two years, since we don't have enough data to link them. Furthermore we 
have the possibility to use a long run (3 months) performed in visual band for another project on the same star (Baran et al. 2010). From Figure \ref{vband} we see that the O-C diagram of the V band monitoring shows a good phase coherence over this time scale. 
The uncertainies are the formal errors of the sinusoidal fit, according to Montgomery and O'Donoghue 1999.

\begin{figure}
\includegraphics[width=80mm]{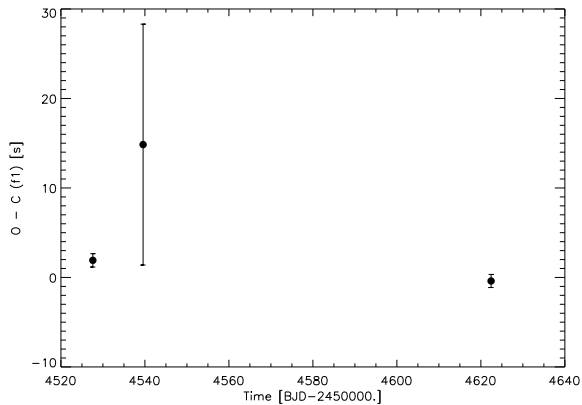}
\caption{O-C diagram of the main frequency (f1) of PG 1325+101 for season 2008 (without V band run).}
\label{2008}
\end{figure}

More difficulties are noticed when we calculate the O-C for the B-band data, probably related to a poor coverage.
Figure \ref{2008} shows preliminary results for the three B-band runs of 2008. 
Considering that the second point has a very large error bar, the phase appears still rather stable.
However, things are more complicate when we join B and V data and we are still working to understand what is the reason of these difficulties.

Similar problems are found also with the data of 2009 and preliminary results are under analysis.
The present coverage in B band is not yet enough to ensure a unique solution, even over a single season.
New data will help to converge towards a unique solution joining together the data of the different seasons with an improved measurement of the pulsation periods.
\\

{\itshape S.B and R.S. acknowledge support from HELAS for participating to the 4th HELAS International Conference in Lanzarote in February 2010.
}

\newpage



\end{document}